\documentclass[reprint,
superscriptaddress,
 amsmath,amssymb,
 aps,
pre,
]{revtex4-2}

\usepackage[usenames,dvipsnames]{xcolor}
\usepackage{xcolor}
\usepackage{hyperref}
\hypersetup{
colorlinks=true,
citecolor=NavyBlue,
linkcolor=NavyBlue,
filecolor=NavyBlue,
urlcolor=NavyBlue
}

\usepackage{comment}
\usepackage{graphicx}
\usepackage{dcolumn}
\usepackage{bm}
\usepackage{orcidlink}
\usepackage[normalem]{ulem}
\usepackage{xcolor}
\usepackage[T1]{fontenc}

\begin{document}

\title{Time-delay estimation using the Wigner-Ville distribution} 
\author{L. de A. Gurgel}
\author{J. M. de Araújo\orcidlink{0000-0001-8462-4280}}
\author{L. D. Machado\orcidlink{0000-0003-1221-4228}}
\author{P. D. S. de Lima\orcidlink{0000-0002-7353-536X}}
\email{paulo.douglas.lima@fisica.ufrn.br}
\affiliation{Departamento de Física Teórica e Experimental, \\Universidade Federal do Rio Grande do Norte, 59078-970 Natal-RN, Brazil}

\date{\today}

\begin{abstract}

Accurately calculating time delays between signals is pivotal in many modern physics applications. One approach to estimating these delays is computing the cross-spectrum in the time–frequency domain. Linear time-frequency representations, such as the continuous wavelet transform (CWT), are widely used to construct these cross-spectra. However, it is well known that the frequency resolution is inherently limited by the localized nature of the convolving wavelet. Moreover, the functional form of the CWT cross-spectrum is not a proper correlation measure and typically requires post-processing smoothing. Conversely, quadratic representations achieve joint time–frequency resolution approaching the Gabor–Heisenberg limit while also providing an adequate measure of similarity between the signals. Motivated by these advantages, we propose a time-delay estimation method based on the Wigner-Ville Distribution (WVD). Considering nonstationary signals arising from two typical wave-physics scenarios, we show that the WVD yields more accurate time-delay estimates with lower uncertainty, particularly in the most energetic frequency bands.
\end{abstract}
\maketitle

\section{Introduction}

Time-delay estimation is a technique for determining differences in the arrival times of signals recorded by different sensors, and it is essential across diverse scientific areas~\cite{time_delay_intro1, time_delay_intro2, time_delay_intro3, time_delay_intro4, giovannelli2025}. For example, it is used as a proxy to detect and denoise gravitational wave signals from space-based detectors~\cite{wvd_gr, time_delay_application1, BADARACCO2024101707}, or to estimate fractal dimensions from chaotic time series~\cite{louis2006}. In dynamical systems, time-delays arise whenever a phenomenon involves transport over a finite distance at finite velocity, leading to an inherent temporal lag between cause and effect~\cite{time_delay_app1, time_delay_app2, time_delay_app3}. In seismology, the time delay between seismic signals encodes information about the Earth's medium and can be used to infer changes in its heterogeneity~\cite {mao2019_2, BrJG2174}. Methods of inverse modeling based on time delay computations are known to mitigate cycle-skipping caused by phase mismatch~\cite{operto2009, warner2016}, and yield more accurate estimates of the medium’s physical properties when weaker prior information is available~\cite {hale_ma_2013, malcolm_2014, charles_2024}. 

Various linear methods for time-delay estimation have been proposed in the literature. The stretching method~\cite{weaver2003, streching_2006, streching_2009}, which assumes that the delay between two signals can be inferred by a series of linear contractions/dilations in one of them. This feature is limited in its applicability over extended time intervals, where nonlinear effects due to media heterogeneity are unavoidable. Another method consists of maximizing the cross-correlation over windowed versions of both signals, either in the time or frequency domain~\cite{coherence_analysis}. In the latter case, it is also referred to as coherence analysis~\cite{MAHARAJ20103516}, since the correct time delay is the one that maximizes the coherence between the signals. This approach is sensitive to the window length, and an inappropriate choice may hinder the detection of cycle-skipping effects. To overcome the window limitation, dynamic time warping (DTW)~\cite{hale2013, matthew2015} has been introduced as an alternative to time-delay estimation. DTW is a nonlinear method that does not assume homogeneity of the underlying medium and can provide accurate time delays even at low signal-to-noise ratios. However, it still fails to provide a complete time-frequency time-delay map, as it retains only temporal information. Moreover, due to the absence of spectral content, time delays estimated via DTW lack key details for physical interpretation, thereby limiting their applicability in wave-physical analysis. Therefore, DTW does not inherently respect physical constraints and can eventually align two unrelated segments.


Cross-spectral methods in the time-frequency domain~\cite{cross_spec_tf, mao2019} have emerged as an alternative tool for obtaining more informative time delays between nonstationary signals. Since nonstationarity is ubiquitous in wave phenomena (multiple scattering, frequency-dependent attenuation, and mode conversion), time–frequency analysis is appropriate for extracting signal features that may vary over time. Most of these methods are formulated as linear time-frequency representations, in which the classical Fourier transform is applied along the frequency axis, considering different time windows. Another class of time-frequency methods relies on quadratic representations, which are based on the signal's energy distribution in the time-frequency plane. Such representations provide a theoretical resolution superior to that of linear representations in the Gabor-Heisenberg limit. In this context, the Wigner-Ville distribution (WVD) is the most commonly used quadratic representation.


The WVD arises in quantum mechanics as a quasiprobability density function for representing particles in phase space~\cite{quantum1}. Although it has been used as a time-frequency tool in wave dynamics problems~\cite{yen1987, safian2006, coste2008, COSTE2010433}, it has never been employed to estimate time delays in nonstationary waves. Motivated by the mathematical properties of these representations, we investigate the WVD's ability to estimate time delays from time-frequency-dependent cross-spectral computation. The remainder of this paper is organized as follows. We describe a general method for time-delay estimation based on time-frequency representations in Sec. \ref{sec:tfa}, and we also review the mathematical and numerical aspects regarding the continuous wavelet transform (CWT) and the WVD. Sec. \ref{sec:results} is devoted to comparing the accuracy of these methods in estimating time delays in two numerical examples concerning wave propagation in complex media, followed by our conclusions in Sec. ~\ref{sec:conc}. 

\section{Time-Frequency Time-Delay Estimation}
\label{sec:tfa}

In this section, we describe the mathematical aspects of the CWT and WVD and their application to time-delay estimation. We assume that the time series of interest are always square-integrable, that is, $x(t), y(t) \in \mathcal{L}^2(\mathbb{R})$.

\subsection{Continuous Wavelet Transform (CWT)}

Since linear time-frequency representations are not unique, different mappings have been proposed in the literature. While the Gabor Transform~\cite{gabor1946, cohen1995} fixes the window size, the continuous wavelet transform (CWT) provides a better multi-resolution tool by changing the window size along the time-frequency domain~\cite{addison2018, arts2022, ventosa}. This adaptive resolution is achieved by convolving the signal with daughter wavelets $\psi_a(t) = \psi\left(t/a\right)/\sqrt{a}$. These wavelets are time-localized zero average functions, constructed from a mother wavelet $\psi(t)$ and deformed by a positive dilatation parameter $a$. The dilation factor is used to increase the frequency (time) resolution at the price of decreasing the time (frequency) resolution, such that the Gabor-Heisenberg uncertainty relation is respected~\citep {mallat}.

With that in mind, the CWT is defined as:
\begin{equation}
	\text{CWT}_x(t, a) = \frac{1}{\sqrt{a}}\int_{-\infty}^{\infty}x(\tau)\psi^*((\tau - t)/a)\,d\tau\,, \label{eq:cwt}
\end{equation}
where the factor $\sqrt{a}$ ensures that the energy is conserved between wavelets, regardless of the value of $a$. The maps computed from Eq. \eqref{eq:cwt} can be converted to the time-frequency domain using the pseudo-frequency representation~\citep{pseudo_freq}. In our analysis, we consider the complex Morlet function~\cite{morlet_a, morlet_b} as the mother wavelet $\psi(t) = \pi^{-1/4}e^{-i\omega_0 t}e^{-t^2/2}$, where $\omega_0$ is the central frequency used to vary the frequency localization.  

Since the CWT given in Eq. \eqref{eq:cwt} stands as a complex representation of the signal, mismatches between $x(t)$ and $y(t)$ lead to the following time–frequency phase shift:
\begin{equation}
    \Phi_{xy}^{\text{CWT}}(t, f) = \arg{\left(\widehat{\text{CWT}_{xy}}(t, f)\right)}\,,
\end{equation}
where the normalized cross-spectrum $\widehat{\text{CWT}_{xy}}(t, f)$ is defined as:
\begin{equation}
    \widehat{\text{CWT}_{xy}}(t, f) = \frac{\text{CWT}_{xy}(t, f)}{\sqrt{|\text{CWT}_{xx}(t, f)|^2|\text{CWT}_{yy}(t, f)|^2}}\,, \label{eq:cross_cwt} 
\end{equation}
with $\text{CWT}_{xy}(t, f) = \text{CWT}_x(t, f)\text{CWT}^*_y(t, f)$. We mention that Eq. \eqref{eq:cross_cwt} can also be used to compute the time-frequency coherence in the CWT framework. However, since it is not a proper coherence measure~\cite{LIU1994151, torrence1999, KANG2019120888, NAEEM2020124235}, a smoothing operator in both time and frequency domains is typically applied to reduce the wavelet influence in the time-delay estimation. By doing so, we avoid coherence values being identically equal to one or zero across all times and frequencies.

From these considerations, the CWT time-delay is finally given by:
\begin{equation}
    \delta t^{\text{CWT}}(t, f) = -\Phi_{xy}^{\text{CWT}}(t, f)/2\pi f\,, \label{eq:time_delay_cwt}
\end{equation}
where the minus sign in Eq. \eqref{eq:time_delay_cwt} comes from the choice of the conjugate term in the definition of the cross-spectrum in Eq. \eqref{eq:cross_cwt}.

\subsection{Wigner-Ville Distribution (WVD)}
\label{sec:wvd}

In contrast to the linear CWT, we now estimate the time delay using a quadratic time-frequency representation. Among various signal representations with an energetic interpretation, the Wigner-Ville Distribution (WVD) possesses desirable properties, including marginality, unitarity, and frequency-support conservation~\cite{Wang_book}. In this regard, the WVD is defined as:
\begin{equation}
\mathrm{WVD}_{xx}(t,f) = \int_{-\infty}^{\infty}
z\left(t + \tau/2\right)
z^{*}\left(t - \tau/2\right)
e^{-i 2\pi f \tau}\, \mathrm{d}\tau\,, \label{eq:wvd}
\end{equation}
where $z(t)$ is the analytical version of $x(t)$, calculated using the Hilbert transform. Equation \eqref{eq:wvd} is also referred to as the auto-WVD since its kernel corresponds to the signal autocorrelation evaluated at time instants parametrized by the time lag $\tau$. This definition gives the WVD a natural correlative interpretation, and therefore, a cross-spectrum calculated from the WVD is valuable when the goal is to extract information about two distinct signals, as in time-delay estimation.

In a similar manner, the time delay computed with the WVD is given by:
\begin{equation}
    \delta t^{\text{WVD}}(t, f) = -\Phi_{xy}^{\text{WVD}}(t, f)/2\pi f\,, \label{eq:time_delay_wvd}
\end{equation}
where the phase shift in this case assumes the following form:
\begin{equation}
    \Phi_{xy}^{\text{WVD}}(t, f) = \arg{\left[\frac{\text{WVD}_{xy}(t, f)}{\sqrt{\text{WVD}_{xx}(t, f)\text{WVD}_{yy}(t, f)
    }}\right]}\,,
\end{equation}
which involves calculating the WVD only three times, rather than six, compared to the CWT case. We stress that the marginalization in the frequency of Eqs. \eqref{eq:time_delay_cwt} and \eqref{eq:time_delay_wvd} provide a time-dependent time-delay function.

\begin{figure}[!htpb]
    \centering   \includegraphics[width=\columnwidth]{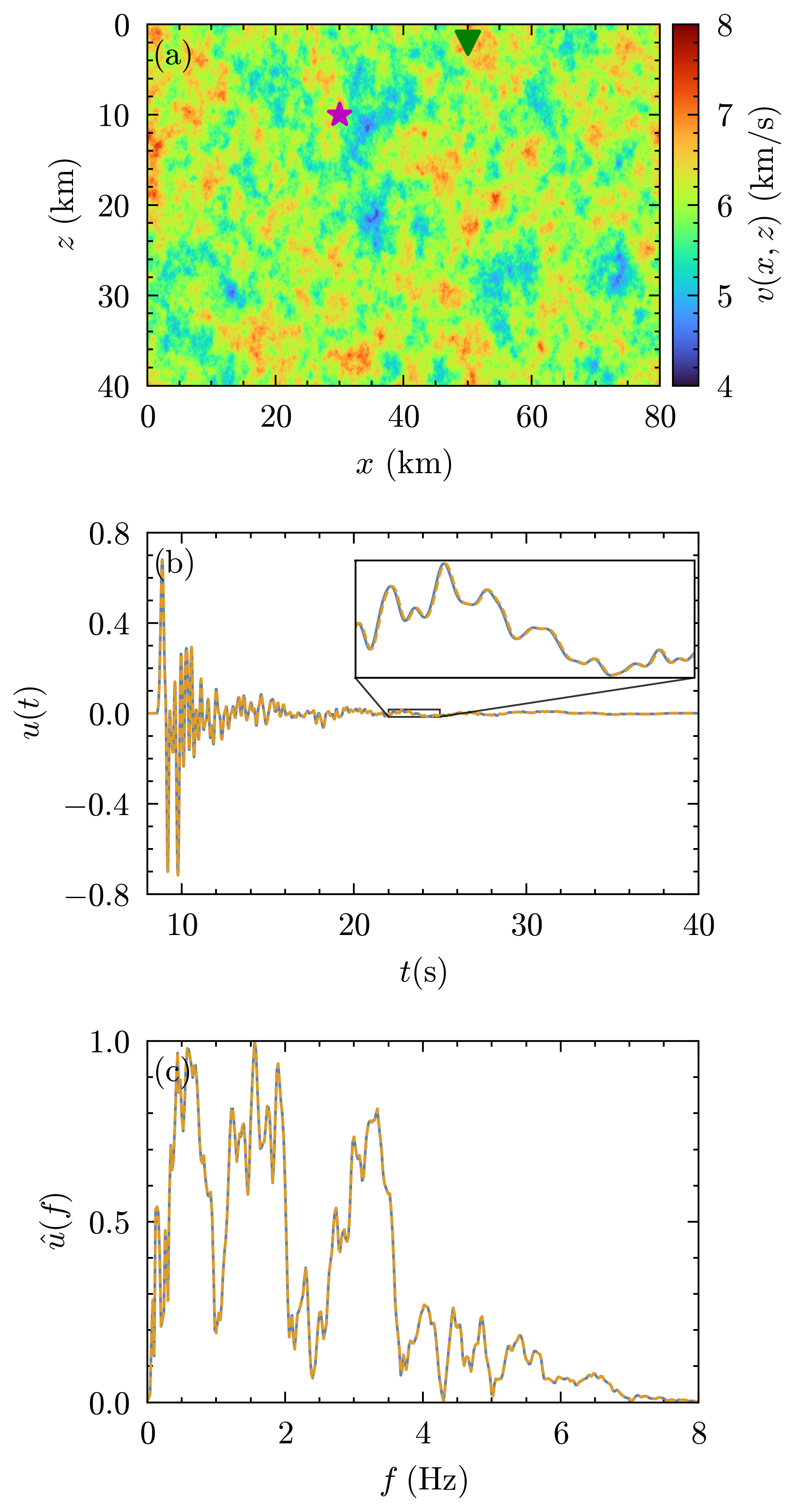}
    \caption{Stochastic medium with spatial correlations considered in the first example. The nonstationary signals (panel b) were recorded by the receiver (green triangle in panel a), which was excited by a source (purple star in panel a). There is no amplitude prior to 8 s. The inset in panel (b) emphasizes that the velocity perturbation in the current medium induces very small time shifts between the signals, as it is also reflected in almost identical normalized frequency spectra (panel c).}
    \label{fig:1}
\end{figure}

A drawback in the WVD for multicomponent signals is the presence of nonphysical cross-terms that arise from the complex conjugate product in Eq. \eqref{eq:wvd}. These undesirable terms overlap with the signal's auto-terms, potentially distorting the WVD spectra. Several strategies~\cite{auto_terms_wvd_1, auto_terms_wvd_2,auto_terms_wvd_3} have been proposed to mitigate this effect. For this work, we analyze the ambiguity domain, which allows explicit separation of cross- and auto-terms.

With this objective, the ambiguity function is defined as:
\begin{equation}
\text{AF}_{xx}(\xi,\tau)
=
\int_{-\infty}^{\infty}
z\!\left(t+\tau/2\right)
z^{*}\left(t-\tau/2\right)
e^{-i 2\pi \xi t}\, dt\,,\label{eq:amb_function}
\end{equation}
which is also easily extendable to a cross ambiguity version $\text{AF}_{xy}(\xi,\tau)$. We also note that the definition of the ambiguity function differs from that of the WVD, since the Fourier transform is taken with respect to the time $t$, rather than the time lag $\tau$.

\begin{figure*}[!htpb]
    \centering
    \includegraphics[width=\linewidth]{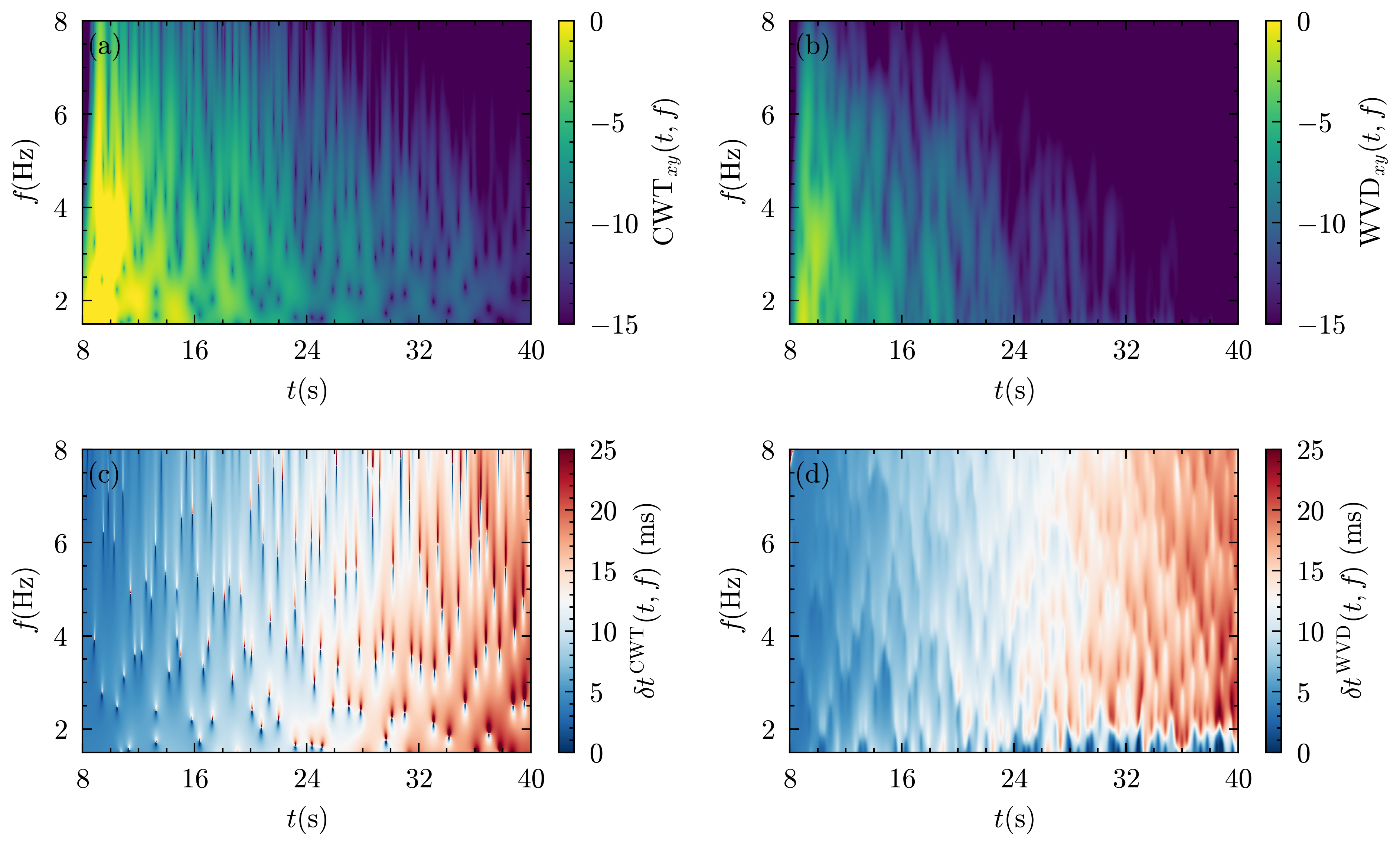}
    \caption{Comparison between the cross-spectrum (shown in log scale) computed using the CWT [Eq. \eqref{eq:cross_cwt}] and WVD [Eq. \eqref{eq:wvd}]. Panels in the bottom row show the corresponding time-frequency time-delay maps obtained with these methods. The central frequency of the Morlet wavelet in the CWT case was $\omega_{0} = 6$ Hz.}
    \label{fig:2}
\end{figure*}

The ambiguity function given by Eq. \eqref{eq:amb_function} has the useful property that the auto-terms are concentrated around the origin $(\xi,\tau) = (0,0)$, while the cross-terms are located far away from it~\cite{Wang_book}. Moreover, whereas the WVD already depends on the time lag $\tau$, the ambiguity function introduces an additional shift $\xi$ along the frequency axis, known as the Doppler frequency~\cite{arne2006}. Therefore, it is necessary to apply a two-dimensional low-pass filter to suppress cross-terms located sufficiently far from the origin in the ambiguity domain. For our time-delay estimation, we first compute three ambiguity functions: $\text{AF}_{xx}$, $\text{AF}_{yy}$, and $\text{AF}_{xy}$. These functions are then filtered to retain only the information in the region dominated by the auto-terms. Finally, the corresponding WVDs are obtained from these filtered analytical signals, producing a time-frequency representation that preserves auto-terms, while effectively mitigating cross-term interference.

\begin{figure}
    \centering
    \includegraphics[width=\columnwidth]{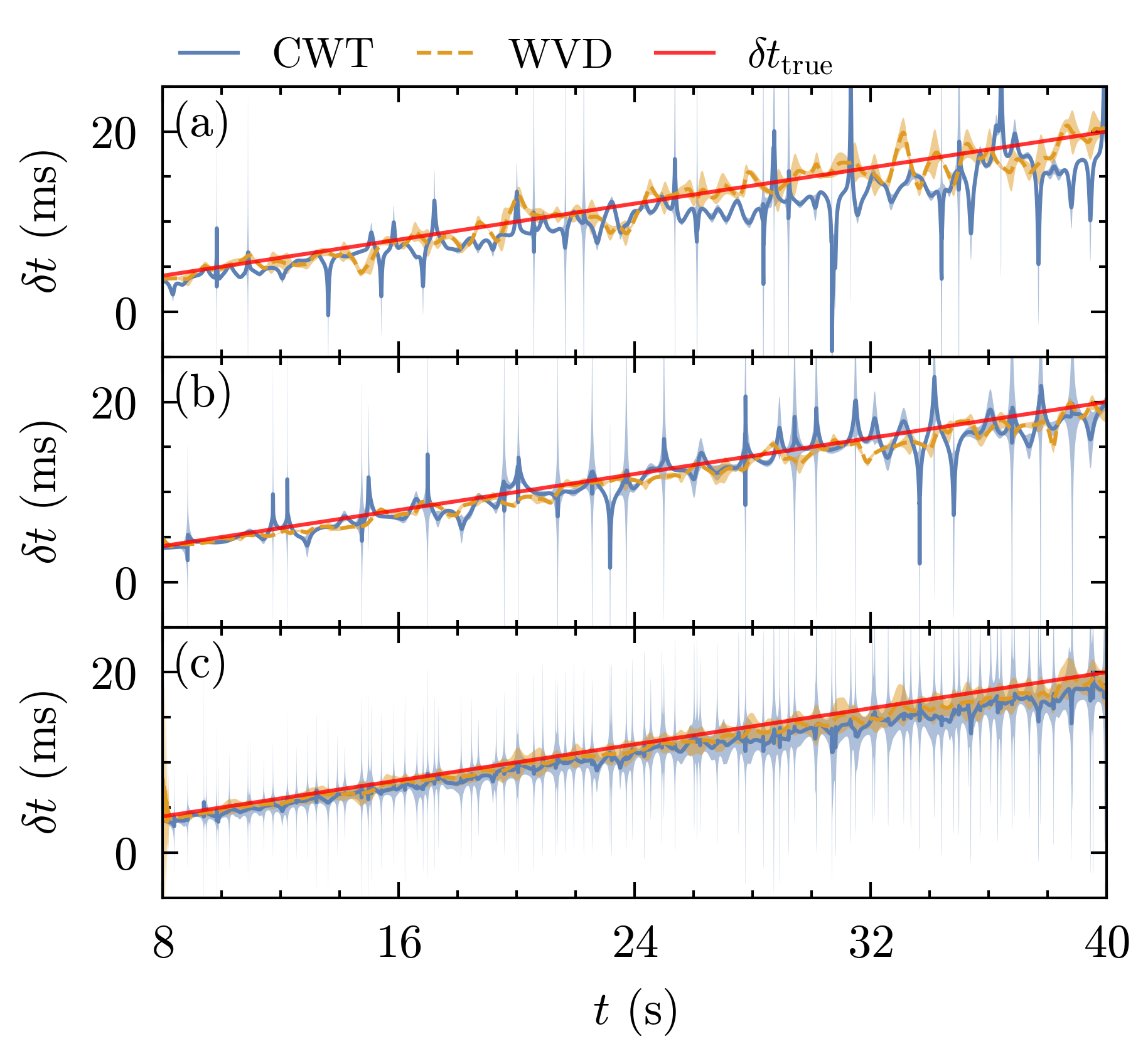}
    \caption{Marginalization of the time-delay maps (see Fig. \ref{fig:2}) for three frequency bands (2.7 - 3.7 Hz, 6.0 - 8.0 Hz, 2.0 - 8.0 Hz). The shaded areas correspond to the standard deviation within each frequency interval. The WVD (orange dashed lines) yields more accurate time delays with lower uncertainty than the CWT (blue solid lines) across these frequency bands.}
    \label{fig:3}
\end{figure}


\section{Results and Discussion}
\label{sec:results}

We now compare the accuracies of the CWT and WVD for estimating signal delays in two distinct wave-physics scenarios. We begin with linear delayed coda waves in a stochastic velocity model, and proceed to an example with a non-linear time delay in a heterogeneous model. These examples are intended to test the proposed method for resolving non-stationary and non-uniform wave distortions.

\subsection{Example 1: Linear Time-Delay in Stochastic Media}
\label{sec:ex1}

For this first example, we consider a two-dimensional medium that possesses coda waves~\cite{coda1, coda2, coda3, coda4}, designed to exhibit strong scattering while avoiding intrinsic attenuation. We disregard elastic effects and use an acoustic approximation for both the reference and current media. In this case, the governing equation to the wavefield $u(\mathbf{x}, t)$ is the constant-density acoustic wave equation $\nabla^2u(\mathbf{x}, t) - v(\mathbf{x})^{-2}\partial^2u(\mathbf{x}, t)/\partial t^2 = f(\mathbf{x}, t)$,
where $v(\mathbf{x})$ is the space-dependent $P$ wave velocity, and $f(\mathbf{x}, t)$ is the source term. Therefore, the time series of interest corresponds to acoustic waves probed by the detector as a response to an excitation source (see green triangle and purple star in Fig. \ref{fig:1}a).

We solve the acoustic wave equation using a finite-difference scheme with sixteen-order spatial derivatives and second-order time derivatives. The model shown in Fig. \ref{fig:1}(a) has 1001 × 501 grid points, corresponding to physical dimensions of 80 km $\times$ 40 km. A damping term~\cite{clayton1977} is included in the wave equation to attenuate wavefields outside the computational boundaries and avoid nonphysical reflection during the simulations. We employ a spatially localized first-derivative Gaussian pulse as the source term, with an amplitude spectrum peaking at approximately $2.5$ Hz.

Figure \ref{fig:1}a illustrates our model, which accounts for heterogeneities that capture roughness at small scales of the Earth~\cite{frankel1988, levander1992, obermann2013}. We introduce spatially correlated perturbations following a von Kármán correlation function in a homogeneous model with a constant velocity of 6000 m/s. The correlation length used is $2000$ m, possessing the same order of magnitude as the dominant wavelength in the background model. The current medium is derived from the reference medium by uniformly increasing the velocity by 0.05\%, ensuring a linear time delay. The waveforms obtained with this procedure are illustrated in Fig. \ref{fig:1}b, where the inset emphasizes very small differences between the signals. This similarity is also reflected in the frequency spectra shown in Fig. \ref{fig:1}c, where the signal energy is concentrated up to 6 Hz. 



The top row of Fig. \ref{fig:2} shows the cross-spectrum computed with CWT and the WVD. The CWT exhibits a stronger cross-spectrum in the time–frequency plane than the WVD, reflecting an overestimation of the underlying wave correlations and, consequently, of the total energy. As an example of this behavior, it is expected that the high-frequency content diminishes in both signals as they evolve due to scattering effects, such that only lower frequencies are present after $\sim$25 s. While the CWT assigns higher energy values to this region, the WVD more accurately captures the lower-energy features, correctly assigning lower cross-spectrum values. Moreover, since the time-delay map (Fig. \ref{fig:2}c) is calculated using the CWT cross-spectrum, this feature is reflected in an underestimation of the time-delays at longer times and higher frequencies, where the signal energy is scarce. This difference between the CWT and WVD arises mainly from the wavelet's intrinsic time–frequency resolution, which leads to a spreading of the cross-spectrum, a phenomenon known as spectral leakage~\cite{mao2019}. 

We attribute the lower resolution in the CWT time-delay maps to the definition of its cross-spectrum. To preserve linearity, the CWT cross-spectrum is formulated as a product of convolutions rather than a true correlation function, as in the quadratic WVD. This fact intensifies the undesirable influence of the wavelet on the time-delay computation, which is typically suppressed by applying a smoothing operator. Accordingly, the main difficulty in using the WVD was determining suitable values for the variables in the ambiguity domain to obtain accurate time delays at frequencies below 2 Hz.

To analyze the difference between the CWT and WVD in more detail, we marginalize the time-delay maps in different frequency bands, as shown in Fig. \ref{fig:3}. These frequency intervals are chosen from the energy distribution of the frequency spectra (see Fig. \ref{fig:1}c). While the first and second frequency bands correspond to the most (Fig. \ref{fig:3}a) and least (Fig. \ref{fig:3}b) energetic portions of the spectrum, the third band (Fig. \ref{fig:3}c) spans the entire frequency range in which the signal exhibits significant energy. Considering the most energetic frequency band (Fig. \ref{fig:3}a), the CWT tends to underestimate the time delay at long times, whereas the WVD follows the correct trend. For the least energetic band, both methods have difficulties dealing with lower amplitudes, but the resolution gain of the WVD prevails, such that the time-delay measurement is much less oscillatory. When the entire spectrum is considered (Fig. \ref{fig:3}c), these features superimpose such that the WVD provides more accurate time-delays than the CWT for this example, regardless of the time duration and frequency content of the wave propagation.

\begin{figure}[!htpb]
    \centering   \includegraphics[width=\columnwidth]{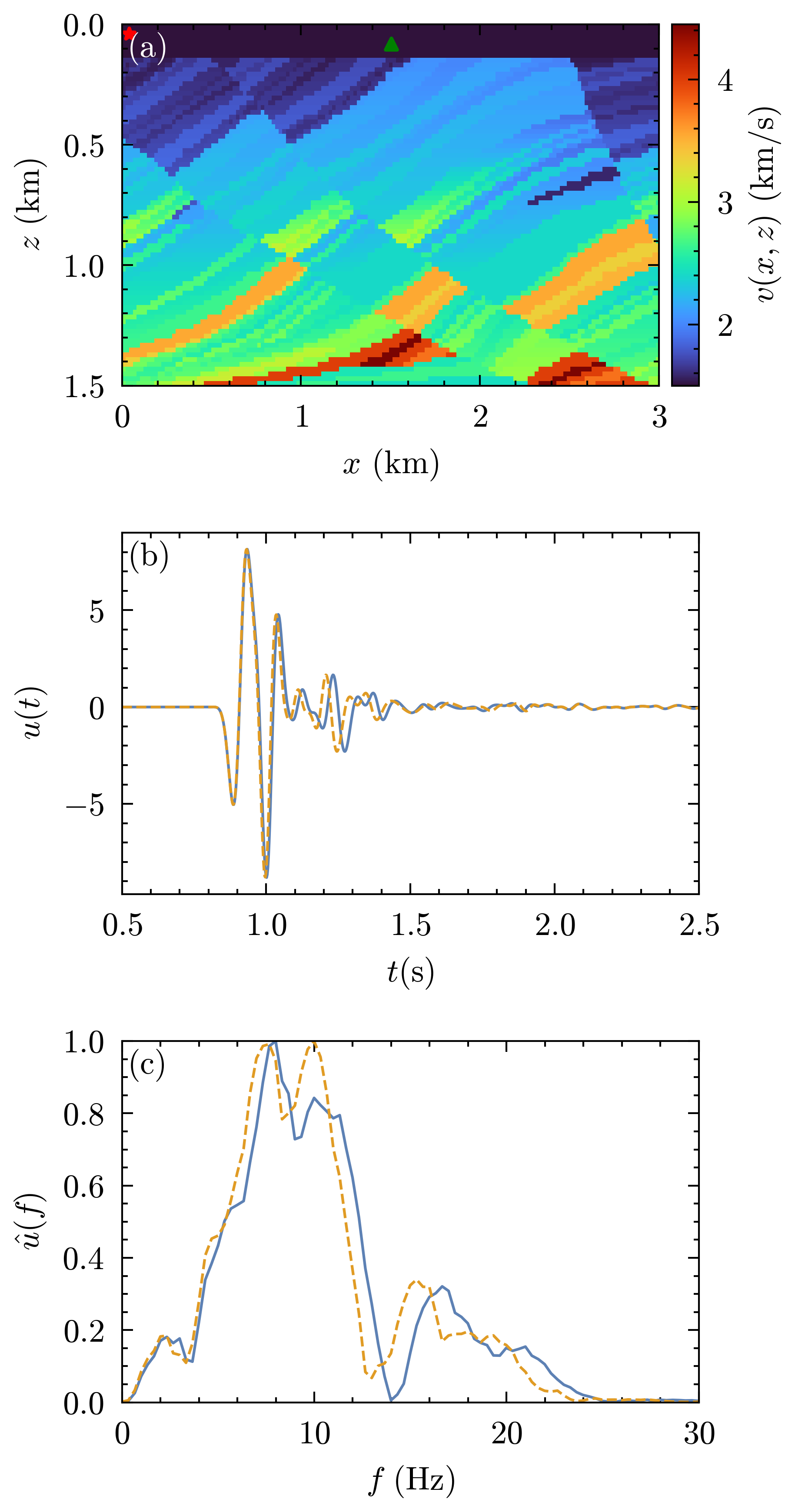}
    \caption{A cropped version of the Marmousi velocity model (panel a) used in the second example. The surface acquisition geometry is represented by a source (red star) and a receiver (green triangle). The current signal is constructed from the reference, including a nonlinear time delay (panel b). This nonlinear time delay induces a pronounced difference in the corresponding frequency spectra (panel c).}
    \label{fig:5}
\end{figure}

\begin{figure*}[!htpb]
    \centering
    \includegraphics[width=\textwidth]{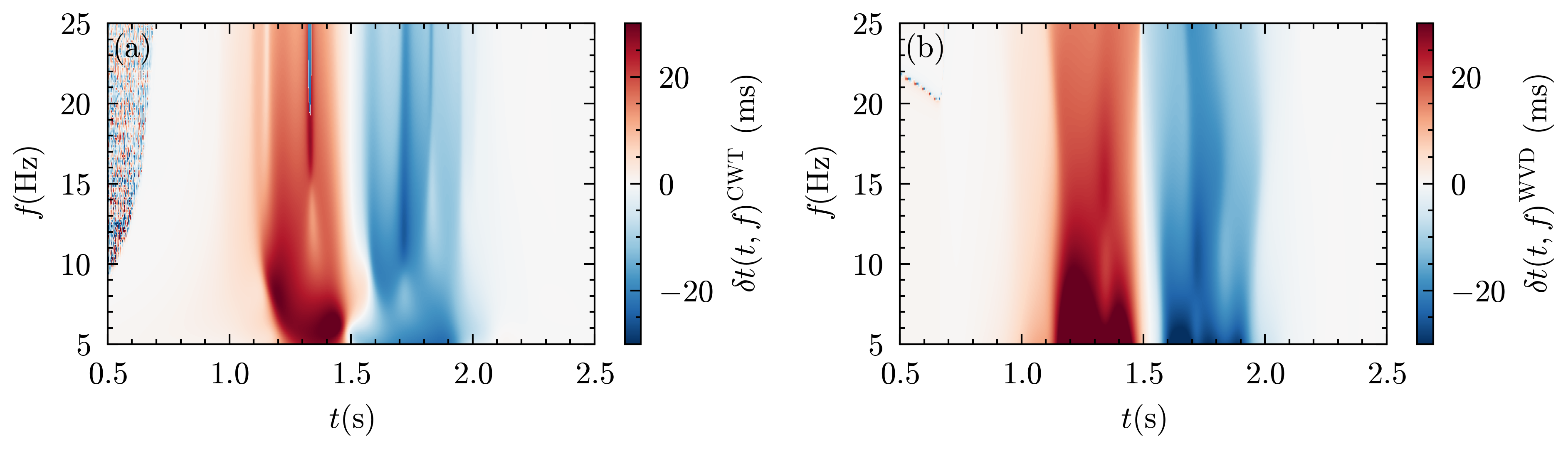}
    \caption{Time-delay maps estimated with the CWT (panel a) and WVD (panel b). The CWT was computed with a wavelet of central frequency $\omega_{0} = 2$ Hz. The WVD more accurately reproduces the trend of the true delay in the most energetic interval and consistently exhibits reduced artifacts.}
    \label{fig:6}
\end{figure*}

\begin{figure}[!htpb]
    \centering
    \includegraphics[width=\columnwidth]{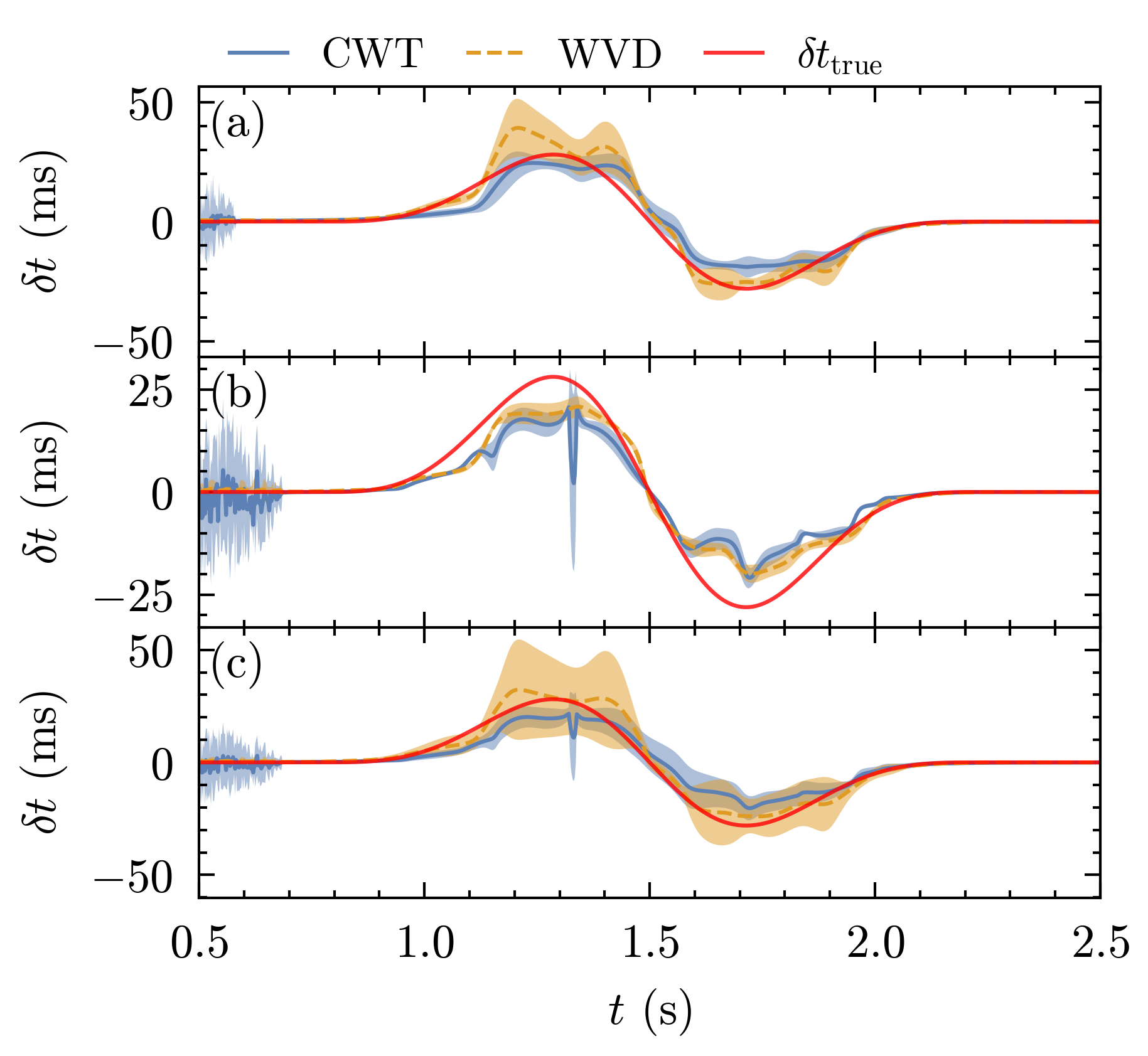}
    \caption{Marginalization of the time-delay maps (see Fig. \ref{fig:6}) for three frequency bands (4 - 12 Hz, 12 - 25 Hz, 2 - 25 Hz). The shaded areas correspond to the standard deviation within each frequency interval. The WVD (dashed orange lines) provides more accurate time-delay estimates within the most energetic frequency band (panel a). However, as with the CWT (blue solid lines), it underestimates delays in the high-frequency range (panel b). Each panel has different maximum and minimum time-delay bounds to clarify the visualization.}
    \label{fig:7}
\end{figure}

An additional advantage of employing time–frequency representations for time-delay estimation is the ability to quantify the uncertainty associated with the selected frequency band. In this regard, spectral leakage in the CWT map increases the uncertainty in time delays (see the peaky regions in Fig. \ref{fig:3}), whereas the WVD yields a consistent uncertainty across the three frequency bands.

\subsection{Example 2: Nonlinear Time-Delay in Heterogeneous Media}

In our second example, we estimate time delays in a cropped section of the Marmousi model (Fig. \ref{fig:5}a). This heterogeneous medium is based on the geology of the Kwanza Basin region in Angola~\cite{versteeg1994} and is widely used as a benchmark for seismic inversion~\cite{martin2006}. The velocity distribution in this case is deterministic and increases with depth, from 1500 m/s at 1500 m to 4500 m/s at 4500 m. The source–receiver geometry considered here corresponds to a marine acquisition configuration, with the source positioned at a depth of 20 m and the receiver located on the ocean bottom at a depth of 100 m (see the red star and the green triangle in Fig. \ref{fig:5}a). We continue to consider an acoustic approximation to wave propagation, using the same numerical setup as in Sec. \ref{sec:ex1}, but with a grid spacing of 20 m and a Ricker wavelet source with a dominant frequency of 10 Hz.

Instead of directly changing the velocity as in the previous example, the current signal (orange dashed line in Fig. \ref{fig:5}a) was designed by interpolating the reference signal (blue solid line in Fig. \ref{fig:5}a) between 0.9 and 2.1 s with a Gaussian time function with zero mean and a standard deviation of 11.35 ms. It mimics a nonlinear perturbation in the medium, such that the current signal exhibits oscillating time-delays with maximum and minimum values of $\pm$ 28 ms (see solid red line in Fig. \ref{fig:7}). As shown in Fig. \ref{fig:5}c, this nonlinear time delay induces significant changes in the frequency spectra, which are both bandwidth-limited to 25 Hz.

Figure \ref{fig:6} displays the time-delay maps obtained using the CWT and the WVD. The energy-spreading characteristic of the CWT yields a map (Fig. \ref{fig:6}a) with artificially smoother time delays, implying that the transform has difficulty assigning correct time delays at transitions near 0.9 s. Similarly to the first example, the CWT tends to correlate the signals more than the WVD (Fig. \ref{fig:6}b), retarding the detection of a delay between the signals. Conversely, the WVD can more effectively capture the timing of this transition for most frequencies. Interestingly, both methods capture the correct trend when the transition occurs around 2.1 s, at which point the negative time shift vanishes.


To verify that the smoother effect of the wavelet in the CWT map can incorrectly represent the true delay, we show the marginalized time delays with their associated uncertainties in Fig. \ref{fig:7} for three different frequency ranges, considering the most energetic band (4–12 Hz), the high-frequency range (12–25 Hz), and the full frequency band (2–25 Hz). We first observe that in the most energetic band (Fig. \ref{fig:7}a), the WVD demonstrates superior performance in predicting the trend of the true time delay, particularly in the region of negative time shifts. A perceived mismatch between the true time delay and the marginalized ones occurs when a high frequency band is considered (Fig. \ref{fig:7}b), although both methods are very similar in their predictions. However, the WVD is slightly superior in the absence of artifacts along the time evolution. Remarkably, when the entire frequency band is used (Fig. \ref{fig:7}c), the confidence interval of the WVD always includes the true time delay and does not display artifacts in the zero-amplitude region (between 0.5 s and 0.8 s) as the CWT. 

We finish our discussion emphasizing the impact of the wavelet's central frequency $\omega_0$ used in the CWT calculation. As we demonstrated, the time-delay map (Fig. \ref{fig:6}b) provides a clearer interpretation, at the expense of less accurate marginalized delays. The opposite occurs when the central frequency is halved. In this case, the marginalized delay approaches the WVD, but the time-delay map exhibits poorer resolution. Although they are not shown here for brevity, the marginalized delays follow the same pattern as in Fig. \ref{fig:7}. From this nonlinear example, we conclude that the WVD is particularly advantageous for time-delay estimation when the signal’s most energetic frequency band is well resolved and available for analysis.

\section{Conclusion}
\label{sec:conc}

We have proposed a time-delay estimation approach based on the Wigner–Ville distribution (WVD). Unlike linear time-frequency transforms, it yields a quadratic representation of the signal. The WVD is well-suited for nonstationary time-series analysis because it provides theoretical optimal time–frequency resolution, as it does not rely on wavelet convolution, unlike the well-established CWT, where the wavelet is inherently more localized than the signal itself. We have also shown that the natural correlative interpretation of the WVD yields time-delay maps that are physically consistent with the observed wave dynamics, without the need to employ a user-defined smoothing operator, as in the CWT case. We compared the two time-frequency representations in two distinct wave-physics scenarios. While the first scenario comprises a stochastic random field with linear-delayed waves, the second is a realistic heterogeneous model with a nonlinear time delay. We have shown that the WVD can estimate time delays more stably and accurately when the most energetic frequency bands are considered. Moreover, we observe that the WVD does not require time-frequency balance to achieve optimal time-frequency resolution and outperforms the CWT, particularly in regions of lower energy content, provided that it avoids spectral leakage. These results suggest that our proposed approach may provide an alternative framework for time-delay analysis, with potential applicability to a range of fields beyond wave modeling.


\begin{acknowledgments}

The authors acknowledge Petrobras' support through the “Development of Seismic Inversion Methodologies for 4D Reservoir Monitoring” project at Universidade Federal do Rio Grande do Norte (UFRN), as well as the strategic importance of ANP's support through the R\&D levy regulation. The authors also thank the High-Performance Computing Center (NPAD) at UFRN for providing computational resources.  L.D.M thanks the Brazilian Research Agencies CAPES and CNPq for their financial support.
\end{acknowledgments}

\bibliography{time_delay_wvd}

\end{document}